\documentclass[aoas,MSNbibl,nameyear,dvips]{arximspdf}
\usepackage{dcolumn}
\usepackage{mathrsfs}
\usepackage{graphicx}

\doi{10.1214/13-AOAS629} %kopijuoti is PTS
\volume{7}
\issue{2}
\pubyear{2013}
\firstpage{1192}
\lastpage{1216}

\makeatletter

\newcolumntype{d}[1]{D{.}{.}{#1}}

\newcommand{\cal}{\mathcal}

\def\BB{\mathbb B}
\def\Z{\mathbf{Z}}
\def\Y{\mathbf{Y}}
\def\Psibf{\bolds{\Psi}}
\def\betabf{\bolds{\beta}}

\makeatother

\begin{document}
\begin{frontmatter}

\title{Spatial risk mapping for rare disease with hidden Markov fields
and variational EM}
\runtitle{Spatial risk mapping with HMRF and variational EM}

\begin{aug}
\author[A]{\fnms{Florence} \snm{Forbes}\corref{}\ead[label=e2]{florence.forbes@inrialpes.fr}},
\author[B]{\fnms{Myriam} \snm{Charras-Garrido}\ead[label=e4]{myriam.charras-garrido@clermont.inra.fr}},
\author[A]{\fnms{Lamiae} \snm{Azizi}\ead[label=e1]{lamiae.azizi@inrialpes.fr}},
\author[A]{\fnms{Senan} \snm{Doyle}\ead[label=e3]{senan.doyle@inrialpes.fr}}
\and
\author[B]{\fnms{David} \snm{Abrial}\ead[label=e5]{david.abrial@clermont.inra.fr}}
\runauthor{F. Forbes et al.}
\affiliation{INRIA Grenoble Rh\^one-Alpes and Grenoble University,
INRA, INRIA Grenoble Rh\^one-Alpes, Grenoble University and INRA,
INRIA Grenoble Rh\^one-Alpes and Grenoble University, and INRA}
\address[A]{F. Forbes\\
L. Azizi\\
S. Doyle\\
Equipe Mistis\\
INRIA Grenoble Rh\^one-Alpes\\
Montbonnot\\
655 avenue de l'Europe\\
38334 Saint-Ismier Cedex\\
France\\
\printead{e1}\\
\hphantom{E-mail: }\printead*{e2}\\
\hphantom{E-mail: }\printead*{e3}} %adresu isvedimo komanda gale!
\address[B]{M. Charras-Garrido\\
D. Abrial\\
INRA UR346\\
F-63122 Saint-Gen\`es-Champanelle\\
France\\
\printead{e4}\\
\hphantom{E-mail: }\printead*{e5}}
\end{aug}

% HISTORY:
\received{\smonth{6} \syear{2011}}
\revised{\smonth{1} \syear{2013}}

% ABSTRACT
%
\begin{abstract}
Current risk mapping models for pooled data
focus on the estimated risk for each geographical unit. A risk
classification, that is, grouping of geographical units with
similar risk, is then necessary to easily draw interpretable maps,
with clearly delimited zones in which protection measures can be
applied. As an illustration, we focus on the Bovine Spongiform
Encephalopathy (BSE) disease that threatened the bovine production
in Europe and generated drastic cow culling. This example features
typical animal disease risk analysis issues with very low risk
values, small numbers of observed cases and population sizes that
increase the difficulty of an automatic classification.
%geographical units into risk classes which are intrinsically not
%so well separated.
We propose to handle this task in a spatial clustering framework
using a nonstandard discrete hidden Markov model prior designed
to favor a smooth risk variation. The model parameters are
estimated using an EM algorithm and a mean field approximation for
which we develop a new initialization strategy appropriate for
spatial Poisson mixtures. Using both simulated and our BSE data,
we show that our strategy performs well in dealing with low
population sizes and accurately determines high risk regions, both
in terms of localization and risk level estimation.
\end{abstract}

% KEYWORDS

\end{frontmatter}

%s1 #&#
\section{Introduction}
The analysis of the geographical variations of a disease and their
representation on a map is an important step in epidemiology. The
goal is to identify homogeneous regions in terms of disease risk
and to gain better insights into the mechanisms
underlying the spread of the disease.
It has long ago become clear that spatial dependencies had to be
taken into account when analyzing such location dependent data.
Most statistical methods for risk mapping of aggregated data
[e.g., \citet{Mollie1999,Richardson1995,Pascutto2000,Lawson2000}]
are based on a Poisson log-linear mixed
model and follow the one proposed by %Besag, York and Molli\'e
\citet{Besag91}. The so-called BYM model introduced %in $1991$
by
\citet{Besag91}, extended by %by Clayton and Bernardinelli in
%$1992$
\citet{Clayton1992} and called the convolution model %by Molli\'e
by
%$1996$
\citet{Mollie1996}, is one
of the most popular approaches and is used extensively
in this context. This model is based on a Hidden Markov Random
Field (HMRF) model where the latent intrinsic risk field is
modeled by a Markov field with continuous state space, namely, a
Gaussian Conditionally Auto-Regressive (CAR) model. In particular,
recent developments in this context concern spatio-temporal
mapping [\citet{Knorr2003,Robertson2010,Lawson2010}] and
multivariate disease mapping [\citet{Knorr2002,MacNab2010}]. For all
these procedures, the
model inference therefore results in a real-valued estimation
of the risk at each location and one of the main reported
limitations [e.g., by \citet{Green2002}] is that local
discontinuities in the risk field are not modeled,
leading to potentially oversmoothed risk maps. Also, in some
cases, coarser representations where areas with similar risk
values are grouped are desirable [e.g., \citet{Abrial2005b}].
Grouped representations have the advantage of providing clearly
delimited areas for different risk levels. %, which is
This can be helpful for decision-makers to
interpret the risk structure and determine
protection measures %
such as reinforced surveillance, movement restriction, %or
mass vaccination or culling (applied in %.In
the Bovine Spongiform Encephalopathy (BSE) example we present). %,
%the cow culling could have been driven by these high risk regions.
From an exploratory point of view, more focused studies could also
be conducted in specific risk regions, in particular, high risk
regions, to better understand the disease determinants.
These areas at risk can be viewed as clusters as in %Knorr-Held and
%Rasser
\citet{Knorr2000}, but we prefer to interpret them as risk
classes, as in the seminal work of
\citet{Schlattmann1993} and \citet{Bohning2000}, and with additional
spatial constraints by \citet{Green2002} and \citet{Alfo2009},
since geographically separated areas can have similar risks and be
grouped in the same class. Consequently, the classes can be less
numerous than the clusters and their interpretation by
decision-makers easier in terms of risk value.
Using the BYM
model, it is possible to derive such a grouping from the output,
using either fixed risk ranges (usually difficult to choose in
practice) or more automated clustering techniques
[e.g., \citet{Fraley2007}]. In any case, this post-processing step
is likely to be suboptimal. In contrast, in this work we
investigate procedures that include such a risk classification.
% lacks a post hoc
%classification of the posterior estimate of the epidemiological
%measure.

There have been several attempts to take into account the presence
of discontinuities in the spatial structure of the risk. Within
hierarchical approaches, one possibility is to move the spatial
dependence one level higher in the hierarchy. %Green and Richardson
\citet{Green2002} proposed replacing the continuous risk field by
a partition model, involving the introduction of a finite number
of risk levels and allocation variables to assign each area under
study to one of these levels. Spatial dependencies are then taken
into account by modeling the allocation variables as a discrete
state-space Markov field, namely, a spatial Potts model. This
results in a discrete HMRF modeling. In the same spirit, in
\citet{Fernandez2002}, the spatial dependence is pushed yet one
level higher. Of course, the higher the spatial dependencies in
the hierarchy, the more flexible the model, but also the more
difficult the parameter estimation. As regards inference, these
various attempts all use
%simulation intensive Monte Carlo Markov Chain
MCMC techniques which can seriously limit, and even prevent, their
application to large data sets in a reasonable time.

Following the idea of using a discrete HMRF model for disease mapping,
we build on the standard hidden Markov field model
used in \citet{Green2002} and \citet{Alfo2009} by considering
a more general
formulation that is able to encode more complex interactions than the
standard Potts model (Section \ref{HMRF}). In particular, we are
able to encode the
fact that risk levels in neighboring
regions cannot be too different, whereas the standard Potts model
penalizes neighboring risks equally, whatever the
amplitude of their difference.

We then (Section \ref{EMalgo}) propose
to use for inference, as an alternative to simulation based techniques,
an Expectation Maximization (EM) framework [\citet{Dempster1977}]
combined with
some variational approximation for tractability in the case of Markov
dependencies.
%In particular, we consider the so-called mean field principle that
%provides
%a deterministic way to deal with intractable Markov Random Field (MRF)
%models
%applications.
An attempt in this direction has been recently made %by Alfo $
by \citet{Alfo2009} but with a rather limited consideration for
experimental validation and robustness of their setting. The
approach in \citet{Alfo2009} has been tested on a single data set
regarding human heart disease and no difficulties regarding
initialization and model selection have been reported.
%This is far from being the case in all practical problems.
In this paper we
% go beyond the work of \citet{Alfo2009}and to
investigate the model behavior in detail. We pay special attention
to one of the main inherent issues when using EM procedures,
namely, algorithm initialization (Section \ref{stra}). We
show that in contrast to the example in \citet{Alfo2009}, simple
initializations do not always work, especially for rare diseases
for which the risks are small, in low population size, as can
occur in animal epidemiology. We then propose and compare
different initialization strategies in order to determine a robust
initialization strategy for most situations that arise in
practice. The model selection issue, for example, the determination
of the number of classes, is addressed using previous work
[\citet{ForbesPeyrard03}] in which a mean field approximation of the
Bayesian Information Criterion (BIC) is provided for HMRF models.
Results are reported %in {\bf Section \ref{EXPE}}
on both simulated data sets (Section
\ref{EXPE}) and a real data set (Section
\ref{esb}) concerning the BSE epidemic in France. The BSE example
is typical of the difficulties that can be encountered. It is a
very rare disease (the global risk in France is about $10^{-4}$)
and concerns a very heterogeneous cow population [see Figure
\ref{cattlehistopop}(a)], where many geographical units have a
very low population (sometimes only one cow), for example, in the
French Riviera. A discussion ends the paper (Section
\ref{DISCU}).

%s2 #&#
\section{Designing hidden Markov fields for spatial disease
mapping} \label{HMRF}

In order to draw interpretable maps,
with clearly delimited zones, %W
we recast the disease mapping issue as a clustering task.
Based on
count data for a rare phenomenon observed in a predefined set $S$
of $N$ areas (e.g., geographical regions), the goal is to
assign to each region a risk level among a finite set of $K$
possible levels $\{\lambda_1,\ldots, \lambda_K\}$ when these risk
levels are themselves unknown and need to be estimated.
% In addition, as already mentioned, we are interested in accounting
%for spatial dependencies between counts in various regions in
%order to get spatially consistent risk estimation.
In general,
risks are expected to be more similar in nearby areas than in
areas that are far apart. The idea is to
exploit the risk information from neighboring areas to provide
more reliable risk estimates in each area.
%We are interested in the modelling of spatial heterogeneity for
%count data on a rare phenomenon, observed in a predefined set of
%areas.
%The domain under consideration is therefore denoted by $S$ and is
%partitioned into $N$ areas.
In each area, two values are usually available, the number $y_{i}$
($i \in S =\{1,\ldots, N\}$) of observed cases of the given
disease and the population size $n_{i}$. A common assumption is
that for an area indexed by $i \in S$, the number of cases $y_i$
is a realization of a Poisson distribution whose parameter depends
on the risk level assigned to the area.
%The unknown risk assignment and risk level compose respectively the
%hidden part and unknown
%parameters of the model. In the following,
It is then convenient to consider the risk assignment for area $i$
as $z_i$ in a set of $K$-dimensional indicator vectors ${\cal L}=
\{e_1,\ldots, e_K\}$, where each $e_k$ has all its components set
to 0 except the $k$th which is 1.

Therefore, the data is naturally divided into observed data
$\mathbf{y} = \{y_1,\ldots, y_N\}$ and unobserved or
missing membership
data $\mathbf{z}= \{z_1,\ldots, z_N\}$. The latter are
considered as
random variables denoted by $\Z= \{Z_1,\ldots, Z_N\}$. The
dependencies between neighboring $Z_i$'s are then modeled by
further assuming that the joint distribution of \{$Z_1,\ldots,
Z_N$\} is a discrete MRF on the graph connecting contiguous
locations (i.e., regions $i$ and $j$ are neighbors if they
are spatially contiguous):
%
%e1 #&#
\begin{equation}\label{pgibbs}
P(\mathbf{z}; \bolds\beta) = W(\bolds\beta)^{-1} \exp\bigl(-H(
\mathbf{z}; \bolds\beta) \bigr),
\end{equation}
where $\bolds\beta$ is a set of parameters, $W(\bolds\beta)$ is a
normalizing constant and $H$ is a function
restricted to pair-wise interactions,
\[
H(\mathbf{z}; \betabf) = - \sum_{i \in S}
z_i^t \alpha- \mathop{\sum_{i,j}}_{i \sim j}
z_i^t \BB z_j,
\]
where we write
$z_i^t$ for the transpose of vector $z_i$ and $ i \sim j $ when
areas $ i $ and $ j $ are neighbors. The set of parameters $
\betabf$ consists of two sets $ \betabf= (\alpha, \BB)$.
Parameter $ \alpha$ is a $K$-dimensional vector which acts as
weights for the different values of $z_i$. When $ \alpha$ is
zero, no risk level is favored, that is, for a given area $ i $,
if no information on the neighboring areas is available, then all
risk levels have the same probability. Then $ \BB$ is a $ K
\times K $ matrix that encodes interactions between the different
classes. If in addition to a null $\alpha$, $ \BB=b \times I_K $
where $ b $ is a real scalar and $ I_K $ is the $ K \times K $
identity matrix, parameters $ \betabf$ reduce to a single scalar
interaction parameter $ b $ and we get the Potts model
traditionally used for image segmentation.
%In this case, parameter
%$ b $ can be interpreted as a strength of interaction between
%neighbors. The higher $ b $, the more weight is given to the
%neighbors. If $ b $ is set to $ 0 $, only the individual features
%are taken into account, reducing our model to the independent non
%spatial case.

Note that the standard Potts model is often
appropriate for classification since it tends to favor neighbors
that are in the same class (i.e., have the same risk level).
However, this model penalizes pairs that have different risk
levels with the same penalty, regardless of the values of these
risk levels. In practice, it may be more appropriate, from a
disease mapping point of view, to encode higher penalties when the
risk levels are further apart. This models the
undesirability of abrupt changes in neighboring risk levels, as it
is unlikely to observe a very low risk area next to a very high
risk area.

In practice, these parameters can
be tuned according to experts, a priori knowledge, or
they can be estimated from the data. In the disease mapping
context, we propose to use for $\BB$ a matrix with three nonzero
diagonals defined for some positive real value $b$ by
%
%e2 #&#
\begin{eqnarray}\label{Bdef}
\BB(k,k) &= & b \qquad\mbox{for all $k=1,\ldots, K$},
\nonumber
\\
\BB(k,l) & = &b/2 \qquad\mbox{for all $(k,l)$ such that $|k-l|=1$},
\\
\BB(k,l) & = &0 \qquad\mbox{otherwise.}
\nonumber
\end{eqnarray}
%
%Equivalently, we can define the above $\BB$ by setting $\BB(k,l) =
%b [1 - |k-l|/2]_+$ where $[ ]_+$ denotes the function equal to
%zero when its argument is not positive and to identity otherwise.
The idea is to favor neighbors in the same risk class first, and
then neighbors in risk classes that are close, with all other
pairs of risk classes being equally weighted. This is the simplest
nonstandard $\BB$ structure that can encode smooth variations in
the risk level. We tested other forms with less
null values in $\BB$ and it
appeared that allowing nonzero entries for pairs of classes
more than 1 risk level apart was not penalizing enough.
More generally, when prior
knowledge indicates that, for example, two given classes are
likely to be next to each other, this can be encoded in the
matrix with a higher entry for this pair. Conversely, when there
is enough information in the data, a full free $\BB$ matrix can be
estimated and will reflect the class structure (i.e., which
class is next to which as indicated by the data) and will then
mainly serve as a regularizing term to encode additional spatial
information. However, a full $\BB$ matrix is not a good idea in
our rare disease situation with poorly separated classes,
considering the potentially ambiguous information contained in the
observations. The fine design of $\BB$ may be important in such a
case.

For the model to
be fully defined, the observation model needs then also to be
specified. Typically in rare disease mapping, a Poisson
distribution is used as the class dependent distribution:
%
%e3 #&#
\begin{equation}\label{poissind}
P(Y_i=y_i |Z_i=z_i; \lambda)
= {\cal P}\bigl(y_i; n_i z_i^t
\lambda\bigr) = \exp\bigl(- n_i z_i^t\lambda
\bigr) \frac{(n_i
z_i^t\lambda)^{y_i}}{y_i!},
\end{equation}
where $n_i$ is the population size in area $i$
and $z_i^t \lambda$ with
$\lambda=[\lambda_1,\ldots, \lambda_K]^t$ is a vectorial notation
that indicates the dependence on the specific value of $z_i$ which
determines the risk level.
%Note that in practice, epidemiologists usually prefer to consider
%relative risks rather than absolute risks. Relative risk
%correspond to the ratio between a local and an overall risk. However,
%in the case of a unique population without any structure, the use
%of relative risk is equivalent to the use of absolute risk.

For the distribution of the observed variables
$\mathbf{y}$ %=\{y_{1},\ldots,y_{N}\}$
given the classification
$\mathbf{z}$, %=\{z_{1},\ldots, z_{N}\}$,
the usual conditional independence assumption leads to $P(\Y=
\mathbf{y}|\Z= \mathbf{z}; \lambda)= \prod_{i
\in S} {\cal P}(y_{i}; n_i
z_i^t\lambda)$. It follows that the conditional probability of the
hidden field $\mathbf{z}$ given the observed field $\mathbf{y}$ is
\[
P(\mathbf{z}| \mathbf{y}; \lambda,\bolds\beta) = W(
\bolds\beta)^{-1} \exp\biggl(-H(\mathbf{z}; \bolds\beta) + \sum
_{i
\in S} \log{\cal P}\bigl(y_{i}; n_i
z_i^t\lambda\bigr) \biggr).
\]
The
parameters of this model are then denoted by $\Psibf= (\lambda,
\alpha, \BB)$.

%s3 #&#
\section{Estimating disease maps using variational EM}
\label{EMalgo}

The question of interest is to recover the unknown
assignment map $\mathbf{z}$. To do so, we consider a Maximum
Posterior
Marginal (MPM) principle consisting of assigning each region $ i $
to the class $ e_k $ that maximizes $ P(Z_i=e_{k} | \mathbf
{y}; \Psibf)
$. In this paper, to deal with the missing data and the spatial
dependence structure, we use the EM algorithm [\citet{Dempster1977}]
with some of the approximations presented in \citet{Celeux2003}
based on the mean field principle.

When the model is a Hidden Markov Model with
parameters $\Psibf$,
there are two difficulties
in evaluating the expectation of
the complete log-likelihood required in the EM algorithm. Both the
normalizing constant
$W(\bolds\beta)$ in (\ref{pgibbs}) and the conditional probabilities
$P(z_i | \mathbf{y}; \Psibf)$ and $P(z_i, z_j |
\mathbf{y}; \Psibf)$ for
$j \sim i$ cannot be computed exactly. The approximate EM we
consider decomposes into an E-step that consists of computing
approximate posteriors denoted by $\tilde{t}_{ik}^{(q)}$ and an
M-step in which the risk updates are available in closed-form:
%
%e4 #&#
\begin{equation}\label{riskup}
\mbox{for all $k$}\qquad \lambda_k^{(q)} = \sum
_{i \in
S} \tilde{t}_{ik}^{(q)} y_i
\Big/\biggl(\sum_{i \in S} n_i y_i
\biggr).
\end{equation}
In contrast, the MRF prior parameters $\bolds\beta$ need to be computed
numerically (see details in Appendix A of the supplemental article
[\citet{supplement}]). Once the parameters are estimated, the area
$i$ is assigned to the class $k$ for which the posterior
probability is the highest.

The likelihood function to be maximized generally possesses many
stationary points of different natures. Consequently, convergence
to the global maximum with the EM algorithm may strongly depend on
the starting parameter values. This is be
particularly true in our context where the discrete distributions
and the low risks increase the estimation difficulty. To
anticipate this initialization issue, as in \citet{Biernacki2002},
we adopt a three stage Search/Run/Select strategy whose goal is to
identify the highest likelihood in a reasonable amount of time:

\textit{Search.} Build a search method for generating $M$
sets of initial parameter values. These sets can be either
generated at random or using some initialization strategy (see Section
\ref{stra}).

\textit{Run.} For each initial position from the search step,
run the variational EM algorithm described in the previous paragraph
until a stopping criterion
is satisfied.

\textit{Select.} Select the set of estimated parameter values
that provides the highest likelihood among the $M$ trials.

We focus below on the Search step and describe the initialization
strategy we propose for a more efficient exploration of the parameter space.

%s4 #&#
\section{A search procedure for initializing EM}\label{stra}

To overcome the sensitivity of the EM algorithm to starting values, different
initialization strategies have been proposed and investigated in
the context of independent Gaussian mixtures
[see, e.g., \citet{McLachlan2000}, Chapter 2, or \citet{Biernacki2002}].
Other initialization
strategies have been investigated in \citet{Karlis2003}
for both Gaussian and Poisson mixtures, leading also to the
conclusion that it was advisable to start from several different
initial values to ensure more reliable results.

Most strategies can be divided into two categories: those based on
initial parameter values and those based on
an initial partition of the data.
For the former, however,
as observed by \citet{Biernacki2004} for the standard nonspatial
Gaussian case, at each iteration of the algorithm, the E and M
steps produce estimated values that are not arbitrary but linked
through some equations. The sequence of such estimates corresponds
then to an EM trajectory in the parameter space. Parameter values
randomly drawn do not necessarily belong to an EM trajectory and
this can result in computationally inefficient strategies, as the
maximum likelihood solution necessarily belongs to one of the
possible EM trajectories.

When using random partitions of the data into $K$ groups, starting
values are obtained by computing the parameter estimates in each
group, here the risk level as the ratio of the observed counts in
the group over the population size of the group. They are by
construction in the EM trajectory space, but they tend to provide
values close to each others and then not to explore the space
efficiently (see Figure 1(d) in the supplemental article
[\citet{supplement}] for an illustration).

In the disease mapping context, the initialization issue has not
really been addressed. In this context, \citet{Alfo2009} use 500
runs of a short-length CEM algorithm before their approximate EM.
They report satisfying results with this initialization procedure,
although it is mentioned in \citet{Biernacki2002} that this choice
is generally not a stable strategy because CEM is actually more
sensitive to the starting value than EM itself. We suspect then
that the example in \citet{Alfo2009} is so that there is no real
initialization problem and any strategy would provide a satisfying
solution. Unfortunately, this is not the case
for the BSE data set under consideration in Section~\ref{esb}, for
which we observed a high sensitivity to initialization.
In this
paper, we address initialization following the EM trajectory-based
idea developed in \citet{Biernacki2004}. Compared to the work in
\citet{Biernacki2004}, our task is complicated by the addition of
a spatial Markov prior whose parameters need also to be
initialized. Our first approach is then to focus on the
initialization of the risks, that is, Poisson distributions
parameters $\lambda$. It is interesting to note that whatever the
model for the spatial prior, an equation similar to that in
\citet{Biernacki2004} can be found that links the $\lambda_k$'s
values. Let $n = \sum_{i \in S} n_i$ be the total population size.
At each iteration $q$, we denote by $n_k^{(q)}$ the quantity
\[
n_k^{(q)} = \biggl(\sum_{i \in S}
\tilde{t}_{ik}^{(q)} n_i \biggr)\big/n,
\]
which can be interpreted as the proportion of the population in
the $k$th risk level. It follows easily that $\sum_{k=1}^K
n_k^{(q)} =1$. Using then equation (\ref{riskup}) for the current
risk level estimations, we get
%
%e5 #&#
\begin{equation}\label{lambdabar}
\sum_{k=1}^K n_k^{(q)}
\lambda_k^{(q)} = \sum_{i
\in S}
y_i\big/n= \bar{\lambda}.
\end{equation}
%
%The later quantity denoted by
$\bar{\lambda}$ can be interpreted as an average risk and has the
property to depend on the observed data only. At each iteration of
the algorithm, the current parameter estimates $\lambda_k^{(q)}$
satisfy this equation. Consequently, all EM trajectories are
included in the space defined by this equation. The idea is then
to produce values for the $\lambda_k$'s by sampling in this space.
A simple way to achieve this is to follow the simulation steps
below:

\textit{Step} 1. Values for the $n_k^{(0)}$'s are first drawn using a
Dirichlet distribution ${\cal D}(\pi,\ldots, \pi)$, with $\pi=1$
for a uniform sampling on the space defined by
$\sum_{k=1}^K n_k^{(0)}=1$.

\textit{Step} 2. Then $k$ is chosen at random in the set $\{1,\ldots, K\}$ and the $\lambda_{l}^{(0)}$'s for $l \not= k$ are
drawn uniformly and without replication
in the sample $\{\frac{y_{1}}{n_1},\ldots, \frac{y_{N}}{n_{N}}\}$.
The last $\lambda_{k}^{(0)}$ is set to verify:
$\lambda_{k}^{(0)}= \bar{\lambda}-\sum_{l \neq k} n_k^{(0)}
\lambda_{l}^{(0)}/n_{k}^{(0)}$. %\label{eqstrat}

The number of initial values generated this way is set by the
user. Note, however, that as in \citet{Biernacki2004} for Gaussian
parameters, the later equation in step 2 does not guarantee that
$\lambda_k^{(0)}$ is strictly positive. If this is not the case,
the simulated sample is discarded and the procedure restarted from
step 1. The proposed strategy is illustrated in
Appendix B of the supplemental article [\citet{supplement}]. We show
its ability to explore the parameter space more efficiently
compared to other standard initializations.

To complete our \textit{Search} procedure, we need
in addition starting positions for the Markov prior parameters
$\bolds\beta= (\alpha, \BB)$. When $\BB$ reduces to some value $b$
[definition~(\ref{Bdef})], our full \textit{Search} procedure
decomposes then in two steps:

\textit{Search} 1. Generate $M$ starting values $\lambda^{(0)}$ using
the two steps above.

\textit{Search} 2. For each initial value $\lambda^{(0)}$, set
$\alpha^{(0)}= 0$ (equal class proportions), $b^{(0)}=1$ and run
our variational EM until the chosen stopping criterion is
satisfied,
with $b$ kept fixed to its initial value. Only $\lambda$ and $\alpha$
are updated.
We propose to use a stopping criterion based on a relative change
in log-likelihood.

The idea in adding the second step is to prevent undesirable
behavior of the algorithm in the case of complex or very noisy
data (see Figure 2 in the supplemental article
[\citet{supplement}]). We observed in our real data set (see
Section \ref{esb}) that imposing a certain amount of spatial
structure first could help to avoid converging to meaningless
solutions. This is typically done by fixing $b= b^{(0)}= 1$ for a
number of iterations before letting all the parameters free. This
strategy is a simple solution we found to deal with very low risk
values and poorly separated Poisson mixture
components, as can be observed in epidemiology.
In practice, $b$ needs not to be fixed exactly to 1. The
appropriate range for $b$ depends on the number of neighbors in
the underlying spatial structure. For 4 to 8 neighbors, values
around 1 have been observed, mainly in the 2D image segmentation
context, to lead to reasonable spatial interactions. In our
context, we observe that the $\alpha$ values can become very small,
probably when the model is trying to \textit{remove} classes
difficult to distinguish using the observed data. The model then
seems to compensate for a small $\alpha$ by increasing $b$. Fixing
$b$ for some iterations at first is a way to favor reasonable
estimations of the $\alpha$'s, which in turn prevent an
overestimation of $b$. In a fully Bayesian approach, another way
to prevent overestimation would be to use an exponential prior on
$b$. This choice is convenient in that it does not change much the
numerical procedure used to estimate $b$, but the choice of the
hyperparameter also has to be fixed. Note that in simpler better
separated cases, the Search~2 step is usually not necessary.

%s5 #&#
\section{Illustrations on simulated data sets}
\label{EXPE} 
\begin{figure}

\includegraphics{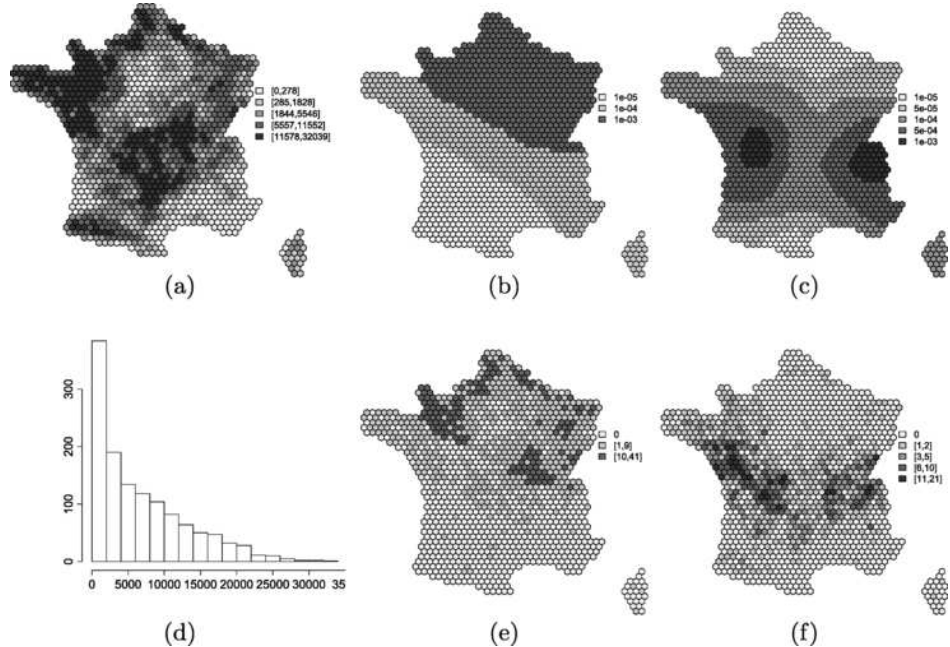}

\caption{Simulated data sets. \textup{(a)}, \textup{(d)}: map and histogram of cattle
population sizes. \textup{(b)},~\textup{(c)}: synthetic underlying risk maps and \textup{(e)}, \textup{(f)}:
simulated counts for the 3-class and 5-class
cases.}\label{cattlehistopop}
\end{figure}

Our goal is to address the analysis of typical rare
animal disease data for which the observed cases and the risk
values may be very low, typically less than 10 cases among a small
population size of few hundreds. In our illustrations (both
simulated and real data), the underlying structure is derived from
the French territory. France is divided into $1264$ hexagons each
of width $23$ km ($450$ km$^2$). The neighborhood structure is based
on adjacent hexagons. For each hexagon, the population size $n_i$
is set to the corresponding cattle population in France in the years
2001--2005: the $n_i$'s vary from $1$ to 32,039 [Figure
\ref{cattlehistopop}(a), (d)]. We consider then different
simulated count data $\mathbf{y}$.

%f1 #&#

For comparison we consider three different strategies to provide
an estimation of the unknown parameters and mapping into regions
of homogeneous risk levels. Two of them, denoted, respectively, by
$S_{\mathrm{tra}}$ and $S_{\mathrm{rand}}$, correspond to Search/Run/Select
decompositions as introduced in Section \ref{EMalgo}. A~third one
denoted by $S_{\mathrm{EMM}}$ represents commonly used strategies when
dealing with initialization issues. In particular, it is close to
the strategy used in \citet{Alfo2009}. The only difference is that
in this later work the nonspatial EM is replaced by a nonspatial
CEM (Classification EM). We rather use EM since CEM is known to be
even less stable than EM with respect to initialization
[\citet{Biernacki2002}]. More specifically:

\textit{$S_{\mathrm{tra}}$ strategy}: $M$ initial values for all the
parameters are generated using the EM trajectory properties and
the full search procedure described in Section \ref{stra}. Our
variational EM is then run for each parameter set until
convergence and the parameter values corresponding to the highest
likelihood are selected.

\textit{$S_{\mathrm{rand}}$ strategy}: This strategy differs from the previous
one only in the way the $M$ initial $\lambda$ values are
generated. They are generated at random (uniformly between 0 and
1.5 in our disease mapping context).

\textit{$S_{\mathrm{EMM}}$ strategy}: $M$ initial values for $\lambda$ are
generated uniformly at random, $\alpha$ is initialized to the null
vector and $b$ is fixed to 0 (nonspatial case). The standard EM
algorithm with no spatial interaction is run until convergence for
each parameter set. The estimate parameter values with the highest
likelihood are then selected and used as initial values for our
variational EM with spatial interaction.

Focusing then on the $S_{\mathrm{tra}}$ strategy, we investigate variations
in the hidden MRF part. We compare the interaction model we
propose [equation (\ref{Bdef})] to the standard Potts model ($
\BB=b \times I_K $) and to another variant for which $\BB$ is so
that $\BB(k,l) = b (1- |k-l|/(K-1))$. Note that for $K=2$ the
formula above leads to the standard Potts model and for $K=3$ to
our model. It distinguishes from our choice for $K>3$ and we will
refer to it as the \textit{smooth gradation model}. The idea behind
such a comparison is to show that three nonzero diagonals in the
$\BB$ interaction matrix (our model) are enough to account for
smooth gradation constraints. Comparison with standard independent
mixtures ($\BB=0$) is not reported, but we observed, as expected,
that such nonspatial models did not provide satisfying risk maps.

Regarding variational EM, we investigated
both the so-called \textit{Mean Field} and \textit{Simulated Field}
variants. In contrast to some other studies [\citet{Celeux2003}], we
observed that for the type of data sets under consideration, the
\textit{Mean Field} algorithm was providing better and more stable
results. This is probably due to the fact that this variant tends
to smooth the data more, which is here an advantage to better
recover the spatial structure. In the following sections, results
are then reported only for the \textit{Mean Field} algorithm.

%s5.1 #&#
\subsection{Typical simulated examples}
\begin{figure}

\includegraphics{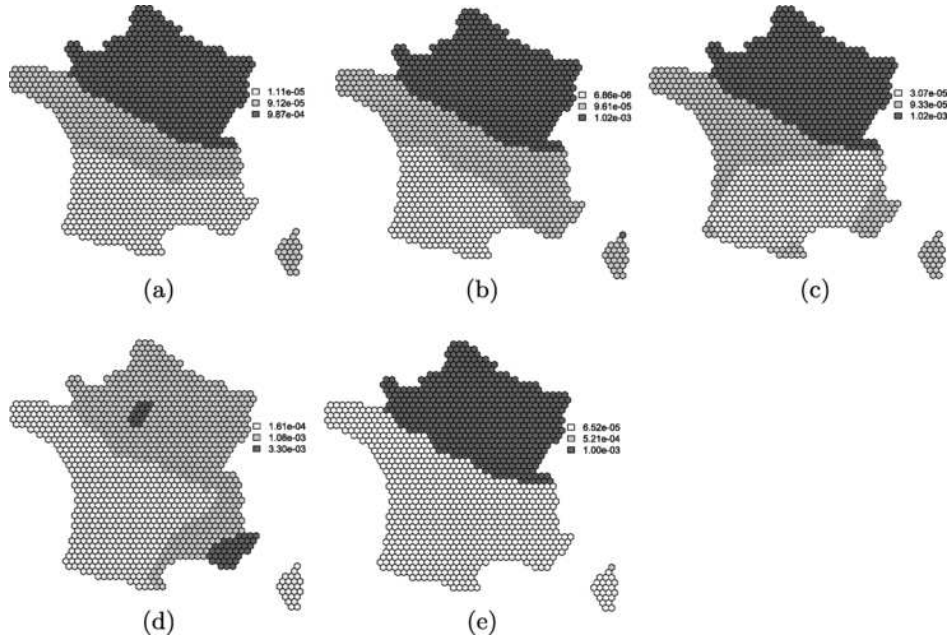}

\caption{Classification results in the 3-class case. \textup{(a)}, \textup{(b)}, \textup{(c)}:
risk maps obtained, respectively, with the $S_{\mathrm{tra}}$, $S_{\mathrm{rand}}$ and
$S_{\mathrm{EMM}}$ strategy. \textup{(d)}: risk map obtained with the BYM model.
\textup{(e)}: risk map obtained with the standard Potts model using
$S_{\mathrm{tra}}$.} \label{res3class}
\end{figure}

\begin{figure}

\includegraphics{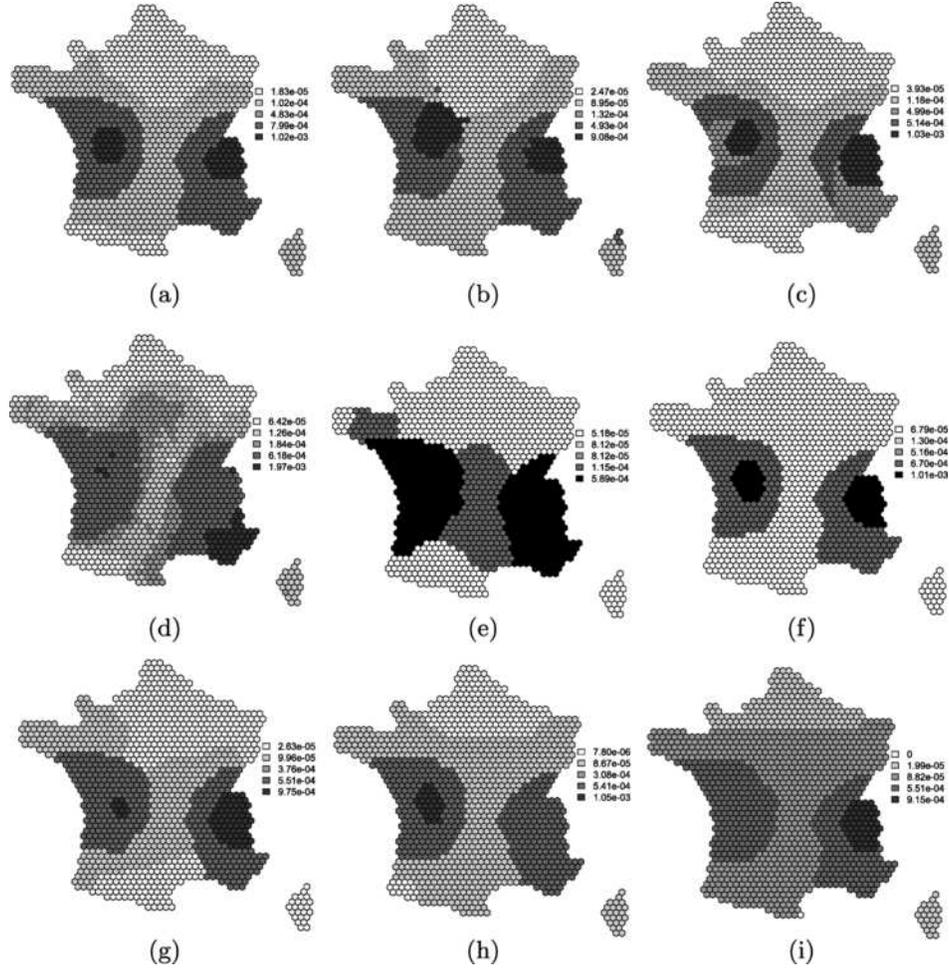}

\caption{Classification results in the 5-class case. Risk maps
obtained with \textup{(a)} $S_{\mathrm{tra}}$, \textup{(b)} $S_{\mathrm{rand}}$,
\textup{(c)}~$S_{\mathrm{EMM}}$,
starting from 1000 initial positions,\textup{(d)} with the BYM model, \textup{(e)}
using the standard Potts model and \textup{(f)} the \textit{smooth gradation
model}. Risk maps obtained starting from 10 initial positions with
the \textup{(g)} $S_{\mathrm{tra}}$, \textup{(h)} $S_{\mathrm{rand}}$ and \textup{(i)} $S_{\mathrm{EMM}}$ strategies.}
\label{res5class}
\end{figure}

We consider two synthetic risk maps with, respectively, $3$ and $5$
risk classes [see Figure \ref{cattlehistopop}(b), (c)]. In the
3-class case, risk levels are set to $\lambda_1= 1\times
10^{-5}$, $\lambda_2= 1\times 10^{-4}$ and $\lambda_3= 1\times
10^{-3}$.
%which correspond in
%epidemiological terms to {low, medium} and {high} risk
%levels.
In the 5-class case, the risks are set to $\lambda_1= 1\times
10^{-5}$, $\lambda_2= 5\times10^{-5}$, $\lambda_3= 1\times
10^{-4}$, $\lambda_4= 5\times10^{-4}$
and $\lambda_5= 1\times10^{-3}$, %.
% corresponding to {\it very low, low,
%medium, high} and {\it very high} risk levels.
corresponding to diseases as rare as BSE. From
the population counts ($n_i$'s), the \textit{true} risk values above,
the known classes, we can easily simulate the counts $y_i$'s from
the Poisson distribution in (\ref{poissind}). Examples of such
counts are shown in Figure \ref{cattlehistopop}(e), (f). Figures
\ref{res3class} and \ref{res5class}(a)--(c) show the corresponding
classifications obtained with the three strategies $S_{\mathrm{tra}}$,
$S_{\mathrm{rand}}$ and $S_{\mathrm{EMM}}$, with $M=1000$ assuming $K=3$ and $K=5$,
respectively. The classification obtained with the BYM method
[\citet{Mollie1991}] and with the two other $\BB$ models mentioned
above are also reported [Figures \ref{res3class}(d), (e) and
\ref{res5class}(d), (e), (f), resp.]. The performance is
evaluated considering both classification performance and risk
value estimation. For classification performance, we consider for
each class the Dice similarity coefficient (DSC) [\citet{dice45}].
This coefficient measures the overlap between a segmentation
result and the ground truth. Denoting by $\mathrm{TP}_k$ the
number\vspace*{1pt} of true positives for class $k$, $\mathrm{FP}_k$ the number
of false positives and $\mathrm{FN}_k$ the number of false
negatives, the DSC is given\vspace*{2pt} by $d_k = \frac{2 \mathrm{TP}_k}{2
\mathrm{TP}_k + \mathrm{FN}_k + \mathrm{FP}_k }$ and $d_k$ takes
its value in $ [0, 1 ]$, where $1$ represents the perfect
agreement. Table \ref{extable1} shows these DSC values and the
estimated risk values.

%f2 #&#

%f3 #&#

%t1 #&#
\begin{table}
\caption{Three-class and five-class data sets. Dice similarity
coefficient (DSC)
and estimated risk for each class, using our model with different
initialization strategies (first 3 rows) and two other $\BB$
models with strategy $S_{\mathrm{tra}}$: the standard Potts model (4th row)
and the Smooth gradation model when different (5th row in the
five-class case)}\label{extable1}
{\fontsize{8.9pt}{10.9pt}\selectfont{
\begin{tabular*}{\tablewidth}{@{\extracolsep{\fill}}lcd{1.2}c@{}}
\hline
\textbf{True risk level} & \textbf{Strategy} & \multicolumn{1}{c@{}}{\textbf{DSC}}
& \multicolumn{1}{c@{}}{\textbf{Estimated risks}} \\
\hline
\multicolumn{4}{@{}c@{}}{\textit{Results for the three-class data
set}}\\[4pt]
\textit{Low} & $S_{\mathrm{rand}}$ & 0.97 & $6.86\times10^{-6}$ \\
$1\times10^{-5}$ & $S_{\mathrm{EMM}}$ & 0.71& $3.07\times10^{-5}$ \\
& $S_{\mathrm{tra}}$ & 0.84 & $1.11\times10^{-5}$ \\
& $S_{\mathrm{tra}}$ (Standard Potts) &0.58 & $6.52\times10^{-5}$ \\
[4pt]
\textit{Medium} & $S_{\mathrm{rand}}$& 0.97 & $9.61\times10^{-5}$\\
$1\times10^{-4}$ & $S_{\mathrm{EMM}}$ & 0.75 & $9.33\times10^{-5}$ \\
& $S_{\mathrm{tra}}$ & 0.86 & $9.12\times10^{-5}$\\
& $S_{\mathrm{tra}}$ (Standard Potts) & 0.00 & $5.21\times10^{-4}$ \\
[4pt]
\textit{High} & $S_{\mathrm{rand}}$ & 0.99 & $1.02\times10^{-3}$\\
$1\times10^{-3}$ & $S_{\mathrm{EMM}}$ & 0.99& $1.02\times10^{-3}$ \\
& $S_{\mathrm{tra}}$ & 1& $9.87\times10^{-4}$ \\
& $S_{\mathrm{tra}}$ (Standard Potts) & 0.99 & $1.00\times10^{-3}$ \\
%True risk level & Strategy & DSC & Estimated risk \\
[6pt]
\multicolumn{4}{@{}c@{}}{\textit{Results for the five-class data
set}}\\[4pt]
\textit{Very low} & $S_{\mathrm{rand}}$ & 0.59 & $2.47\times10^{-5}$
\\
$1\times10^{-5}$ & $S_{\mathrm{EMM}}$ &0.54 & $3.93\times10^{-5}$ \\
& $S_{\mathrm{tra}}$ & 0.62 & $1.83\times10^{-5}$\\
& $S_{\mathrm{tra}}$ (Standard Potts) & 0.44 & $5.18\times10^{-5}$ \\
& $S_{\mathrm{tra}}$ (Smooth gradation) & 0.33 & $6.79\times10^{-5}$
\\
[4pt]
\textit{Low} & $S_{\mathrm{rand}}$& 0.39 & $8.95\times10^{-5}$\\
$5\times10^{-5}$ & $S_{\mathrm{EMM}}$ & 0.05 & $1.18\times10^{-4}$ \\
& $S_{\mathrm{tra}}$ & 0.24 & $1.02\times10^{-4}$ \\
& $S_{\mathrm{tra}}$ (Standard Potts) & 0 & $8.12\times10^{-5}$ \\
& $S_{\mathrm{tra}}$ (Smooth gradation) & 0 & $1.30\times10^{-4}$ \\
[4pt]
\textit{Medium} & $S_{\mathrm{rand}}$ & 0 & $1.32\times10^{-4}$\\
$ 1\times10^{-4}$ & $S_{\mathrm{EMM}}$ & 0.09 & $4.99\times10^{-4}$\\
& $S_{\mathrm{tra}}$ & 0& $4.83\times10^{-4}$\\
& $S_{\mathrm{tra}}$ (Standard Potts) & 0 & $8.12\times10^{-5}$ \\
& $S_{\mathrm{tra}}$ (Smooth gradation) & 0 & $5.16\times10^{-4}$ \\
[4pt]
\textit{High} & $S_{\mathrm{rand}}$ & 0.84 & $4.93\times10^{-4}$ \\
$5\times10^{-4}$ & $S_{\mathrm{EMM}}$ & 0.76& $5.14\times10^{-4}$\\
& $S_{\mathrm{tra}}$ & 0.91 & $7.99\times10^{-4}$ \\
& $S_{\mathrm{tra}}$ (Standard Potts) & 0.03 & $1.15\times10^{-4}$ \\
& $S_{\mathrm{tra}}$ (Smooth gradation) & 0.89 & $6.70\times10^{-4}$
\\
[4pt]
\textit{Very high} & $S_{\mathrm{rand}}$ & 0.72 & $9.08\times10^{-4}$ \\
$1\times10^{-3}$ & $S_{\mathrm{EMM}}$ & 0.87 & $1.03\times10^{-3}$ \\
& $S_{\mathrm{tra}}$ &0.96 & $1.83\times10^{-3}$\\
& $S_{\mathrm{tra}}$ (Standard Potts) & 0.31 & $5.89\times10^{-4}$ \\
& $S_{\mathrm{tra}}$ (Smooth gradation) & 0.92 & $1.01\times10^{-3}$\\
\hline
\end{tabular*}}}
\end{table}

For both the 3 and 5 class examples, the BYM model is clearly not
providing satisfying mappings. In particular, the highest risk regions
are found in regions with very few cattle populations (e.g.,
South--East of France). In terms of risk estimation, BYM tends to
overestimate risk levels, especially the lowest ones. For high risks,
the overestimation is not as large, but the corresponding regions are
not properly identified. Note, however, that this model has not been
originally designed to handle data simulated from a small number of
constant risk values. This type of data clearly favors HMRF-based
models such as ours, although we suspect the BYM limitations mainly
come from its difficulties in handling low population size (see a
similar conclusion in the BSE case).

To consider cases more favorable to the BYM
model and put our model at comparative disadvantage, we also
simulated a data set similar to that in \citet{Green2002}, with a
North--South gradient corresponding to risks smoothly (linearly)
decreasing from North to South from $10^{-3}$ to $10^{-5}$. The
true risk map, the corresponding simulated data and the BIC values
for the Potts and our models are shown, respectively, in Figure 3(a), (b) and
Table 1 of the supplemental article [\citet{supplement}]. The
selected $K$ is 5 for both models. The resulting maps for $K=3,5$
and 7 are then also shown in Figure 3 for the Potts (e), (h), (k), the
BYM (f), (i), (l) and our models (d), (g), (j). The
discrete HMRF models (Potts and ours) show satisfying results with
some better risk estimations and region shapes obtained with our
model [see, e.g., figures (g), (j)]. The BYM model does well in
estimating high risk regions in the North [Figure 3(l)] but
wrongly classifies as high risk the South--East [Figure 3(c), (f), (i), (l)].
%although we observe that for $K>4$
%the classifications become more sensitive to initialization.
This
surprising good behavior of discrete HMRF models has already been
observed by \citet{Green2002} (Section 4.7), where they mention that
the models perform competitively to the BYM model. Similar
conclusions are drawn for
simulations generated
according to the BYM model [see Section 4.7 in \citet{Green2002}].

Going back to our first experiment, in the 3-class
case (Table \ref{extable1}), all strategies give reasonable
results, for the high and medium risk regions, both in terms of
estimation and classification. The main differences are observed
for the low risk region. Our proposed strategy $S_{\mathrm{tra}}$ performs
better than $S_{\mathrm{EMM}}$ at estimating the low risk value. It is also
better, although comparable to the $S_{\mathrm{rand}}$ strategy. In terms of
mapping, $S_{\mathrm{rand}}$ and $S_{\mathrm{tra}}$ clearly outperform $S_{\mathrm{EMM}}$.
The $S_{\mathrm{rand}}$ result looks visually better, but in terms of
classification rates (Table \ref{extable1}) this is the case only
for the low and medium risk regions.

In the 5-class case (Table \ref{extable1}), all strategies have
trouble separating the low and very low risk regions and tend to
lose a class. For the $S_{\mathrm{EMM}}$ strategy we can visualize the 5
classes, but this is due to the division of the true high risk
region into two classes which correspond to almost the same risk
values ($4.99\times 10^{-4}$ and $5.14 \times 10^{-4}$ for high
risk). The low risk region is not better identified in this case
and two classes are separated although they correspond to the same
risk value. We will see in what follows (Table \ref{lbd53class})
that this seems to be a tendency of the $S_{\mathrm{EMM}}$ strategy. In
terms of classification, the $S_{\mathrm{tra}}$ strategy outperforms the
other strategies for the high and very high risk regions, which
correspond to the risk levels of importance in epidemiology.
Indeed, for immediate control purposes, it is
crucial to detect regions where the disease is more developed,
while low risk regions may help afterward to envisage protection
factors by comparing, for instance, the differences in a number of
covariates between these regions and the high risk ones.

%t2 #&#
\begin{table}
\caption{100 five-class and 100 three-class data sets. Mean
and standard deviation of the Dice similarity coefficient (DSC),
mean and standard deviation of the estimated risk value for each
class using different initialization strategies}
\label{lbd53class}
\begin{tabular*}{\tablewidth}{@{\extracolsep{\fill}}lccc@{}}
\hline
\textbf{True risk level} & \textbf{Strategy} & \textbf{DSC} & \textbf{Estimated risks} \\
\hline
\multicolumn{4}{@{}c@{}}{\textit{Results for the three-class data
sets}}\\[4pt]
\textit{Low} & $S_{\mathrm{rand}}$ & 0.84 (0.25) & $1.02\times 10^{-5}
$ $(3.31\times 10^{-6})$ \\
$1\times 10^{-5}$ & $S_{\mathrm{EMM}}$ & 0.53 (0.33) & $4.12\times 10^{-5}
$ $(3.11\times 10^{-6})$\\
& $S_{\mathrm{tra}}$ & 0.79 (0.25) & $1.49\times 10^{-5} $ $(1.48\times
 10^{-5})$\\
[4pt]
\textit{Medium} & $S_{\mathrm{rand}}$& 0.88 (0.20) & $9.82\times 10^{-5}
$ $(6.06\times 10^{-6})$ \\
$1\times 10^{-4}$ & $S_{\mathrm{EMM}}$ & 0.44 (0.41)& $2.19\times 10^{-4}
$ $(2.13\times 10^{-4})$ \\
& $S_{\mathrm{tra}}$ & 0.77 (0.30) & $1.15\times 10^{-4} $ $(6.84\times
10^{-5})$ \\
[4pt]
\textit{High} & $S_{\mathrm{rand}}$ & 0.99 (0.09) &$9.94\times 10^{-4}
$ $(1.71\times 10^{-5})$ \\
$1\times 10^{-3}$ & $S_{\mathrm{EMM}}$ & 0.93 (0.18) &$9.99\times 10^{-4}
$ $(2.57\times 10^{-5})$\\
& $S_{\mathrm{tra}}$ & 0.96 (0.10) & $9.97\times 10^{-4} $ $(1.74\times
10^{-5})$\\
[6pt]
\multicolumn{4}{@{}c@{}}{\textit{Results for the five-class data
sets}}\\
[4pt]
\textit{Very low} & $S_{\mathrm{rand}}$ & 0.42 (0.29) & $2.17\times10^{-5}
$ $(2.15\times10^{-5})$ \\
$1\times10^{-5}$ & $S_{\mathrm{EMM}}$ & 0.36 (0.24) & $2.58\times10^{-5}
$ $(2.98\times10^{-6})$ \\
& $S_{\mathrm{tra}}$ & 0.56 (0.20) &$ 2.07\times10^{-5} $ $(1.53\times
10^{-5})$\\
[4pt]
\textit{Low} & $S_{\mathrm{rand}}$& 0.29 (0.19) & $7.99\times10^{-5}
$ $(7.53\times10^{-5})$ \\
$ 5\times10^{-5}$ & $S_{\mathrm{EMM}}$ &0.22 (0.18) & $5.43\times10^{-4}
$ $(3.49\times10^{-5}) $\\
& $S_{\mathrm{tra}}$ & 0.29 (0.17) & $9.62\times10^{-5} $ $(4.39\times
10^{-5})$ \\
[4pt]
\textit{Medium} & $S_{\mathrm{rand}}$ & 0.38 (0.25) &$1.74 \times10^{-4}
$ $(1.57\times10^{-4})$ \\
$1\times10^{-4}$ & $S_{\mathrm{EMM}}$ & 0.16 (0.21) &$3.03\times10^{-4}
$ $(2.06\times10^{-4})$\\
& $S_{\mathrm{tra}}$ & 0.09 (0.18) & $3.33\times10^{-4} $ $(1.37\times
10^{-4})$\\
[4pt]
\textit{High} & $S_{\mathrm{rand}}$ & 0.51 (0.33) &$4.58\times10^{-4}
$ $(1.97\times10^{-5}) $\\
$5\times10^{-4}$ & $S_{\mathrm{EMM}}$ & 0.55 (0.33)&$5.74\times10^{-4}
$ $(5.86\times10^{-5})$\\
& $S_{\mathrm{tra}}$ &0.66 (0.38) & $5.57\times10^{-4} $ $(1.05\times
10^{-4})$\\
[4pt]
\textit{Very high} & $S_{\mathrm{rand}}$ & 0.44 (0.18) &$8.71\times10^{-4}
$ $(4.27\times10^{-4})$ \\
$1\times10^{-3}$ & $S_{\mathrm{EMM}}$ & 0.65 (0.34) &$9.78\times10^{-4}
$ $(1.76\times10^{-4})$\\
& $S_{\mathrm{tra}}$ & 0.83 (0.17) & $1.05\times10^{-3} $ $(7.66\times
10^{-5})$\\
\hline
\end{tabular*}
\end{table}

To emphasize the difference between the $S_{\mathrm{tra}}$ and $S_{\mathrm{rand}}$
strategies, we consider the same 5-class data set but reduce the
number of starting values to $M=10$. This can typically be
necessary with a limited amount of computational resources. As
mentioned in Section \ref{stra} and illustrated in Figure 1(e)--(g),
in the supplemental article [\citet{supplement}] this
should benefit to our $S_{\mathrm{tra}}$ strategy which is more efficient
in exploring the parameter space and in finding good
initializations. Indeed, we observe more satisfying mapping
results [see Figure \ref{res5class}(g)--(i)] with $S_{\mathrm{tra}}$ than
with $S_{\mathrm{rand}}$ for similar estimations of the different risk
levels. $S_{\mathrm{tra}}$ is clearly better at identifying the very high
risk regions but also in this case the low risk ones. $S_{\mathrm{EMM}}$ is
clearly providing less satisfying results in this case.

Additional maps obtained with the other mentioned $\BB$ models
show that both the \textit{smooth gradation model} and our model
improve over the standard Potts model that does not include
constraints on risk level gradation. In terms of DSC values, our
model outperforms the \textit{smooth gradation model} in the 5-class
example (Table~\ref{extable1}).

To illustrate the robustness of our model to nonsmooth risk level
gradation, we considered two additional
simulations using the same synthetic risk partitions [Figure
\ref{cattlehistopop}(b), (c)] but with permuted risk levels. In the
3-class example, $\lambda_1= 1\times10^{-5}$, $\lambda_2=
1\times10^{-3}$ and $\lambda_3= 1\times10^{-4}$ so that we
locate now the highest risk next to the lowest one. In the 5-class
case, similarly the risk levels are permuted to $\lambda_1=
5\times10^{-5}$, $\lambda_2= 5\times10^{-4}$, $\lambda_3=
1\times10^{-4}$, $\lambda_4= 1\times10^{-5}$ and $\lambda_5=
1\times10^{-3}$. Results obtained with the BYM model and both
the Potts and our models are shown in Figure 4 of the supplemental
article [\citet{supplement}]. The Potts and our model provide
similar results (see also the Dice scores in Table 2 of the
supplemental article [\citet{supplement}]) more satisfying than BYM.
In the 3-class example, the particularity of our model appears
clearly at the border of two of the classes with some wrongly
classified hexagons. Paradoxically, overall the HMRF models
performance is better than in the smooth gradation case. The fact
that neighboring risk levels are now more different induces better
separated classes and makes the classification easier.

%s5.2 #&#
\subsection{Intensive simulation study}
The very low values of
the risk levels induce some difficulty in interpreting the results
and are responsible for some instability. To further investigate
the algorithms, we repeat the simulations above a hundred times
with the same true risk values. For the true values of $K$, the
performance is then evaluated considering both average
classification performance and risk value estimation.
%The
%use of the median instead of the mean is more appropriate as it
%limits the effect of unstable runs that can badly impact the risk
%values when they live in such a low range.
Table \ref{lbd53class} shows for the 3- and 5-class cases, the
mean and standard deviation of the DSC values for the 100
simulated data sets.
%Similarly, Table
It also shows the mean and standard deviation of the estimated
risk values.

%Comment Median values are...
For the 3-class example, the average estimation of the risk values
is in general close to the real parameter values for the three
strategies. However, $S_{\mathrm{EMM}}$ tends to overestimate low risks.
$S_{\mathrm{tra}}$ and $S_{\mathrm{rand}}$ give similar results. In terms of DSC
values, $S_{\mathrm{rand}}$ outperforms $S_{\mathrm{tra}}$ on average and shows
smaller variances. However, the boxplots of Figure 5(a)--(c) in the
supplemental article [\citet{supplement}] show that the median risk
values are very close for both strategies.
In the 5-class case, $S_{\mathrm{tra}}$ provides better average risks for
the high and very high risk values. In terms of DSC values, the
average values are higher and the variances generally lower for
$S_{\mathrm{tra}}$. For estimated risks, it is also the case for medium to
very high risks. The $S_{\mathrm{tra}}$ strategy seems to provide better
and more stable results for higher risk values, which is a
desirable feature in epidemiology. The boxplots in Figure 5(d)--(h)
in the supplemental article [\citet{supplement}] show that this is
generally compensated by a worse estimation of the medium risk
class compared to $S_{\mathrm{rand}}$ [Figure~5(f)].

For the 3-class case, we can notice that the estimator
$\hat{\lambda}$ predicts observations of the parameter $\lambda$
with good accuracy. However, the estimation of parameters
associated to the highest risk region is much more precise than for
lowest risk. For the 5-class case, larger variations are observed
for the class which disappears in general ($\lambda_3$). The
estimation is also more precise for higher risks than for the
lowest ones.

We then also consider the issue of selecting the
right number of classes. In this case $K$ is not fixed. For each
simulation in the 3-class case, we run our algorithm with the
$S_{\mathrm{tra}}$ initialization, for $K=2$ and $K=3$. We observed that
for $K \geq4$, the algorithm systematically loses a class and
these values of $K$ are never selected. We then used the mean
field approximation of the Bayesian Information Criterion (BIC), as
described in \citet{ForbesPeyrard03} for hidden Markov models, to
select $K$ among $K=2$ or 3. For the $S_{\mathrm{tra}}$ strategy, it
follows that among the 100 simulations, $K=3$ was selected 75
times and $K=2$ was selected 25 times. Similarly, in the 5-class
example, we used the approximate BIC to select a value of $K$ from
2 to 7. $K=5$ was selected 46 times, $K=4$ was selected 42 times,
$K=3$ was selected 12 times and $K=2,6,7$ were never selected.
Similar results were observed for the other strategies. It
confirms especially in the 5-class case that the data we have to
deal with do not correspond to an easy well-separated case.

Focusing then again on the $S_{\mathrm{tra}}$ initialization, we compare
the model used in (\ref{Bdef}) with the standard Potts model and
the so-called \textit{smooth gradation model} in terms of
classification rate and risk estimation. The results are shown in
Table \ref{PottsVarTable}.

%t3 #&#
\begin{table}
\caption{100 five-class and 100 three-class data sets. Mean
and standard deviation of the DICE similarity coefficient (DSC),
mean and standard deviation of the estimated risk value for each
class using different $\BB$ models for the $S_{\mathrm{tra}}$
initialization} \label{PottsVarTable}
\begin{tabular*}{\tablewidth}{@{\extracolsep{\fill}}lccc@{}}
\hline
$\BB$ \textbf{model} & \textbf{Risk level} & \textbf{DSC} &
\multicolumn{1}{c@{}}{\textbf{Estimated risks}} \\
\hline
\multicolumn{4}{@{}c@{}}{\textit{Results for the three-class data
sets}}\\[4pt]
\textit{Standard Potts} & Low & 0.40 (0.34) & $3.35\times
10^{-5} $ $(3.34\times10^{-5})$ \\
$\BB= b I_K$ & Medium & 0.43 (0.37) & $2.17\times10^{-4}
$ $(2.46\times10^{-4})$\\
& High & 0.92 (0.22) & $9.84\times10^{-4} $ $(2.25\times
10^{-4})$\\
[4pt]
\textit{Smooth gradation} & Low & 0.79 (0.25) & $1.49\times
10^{-5} $ $(1.48\times10^{-5})$ \\
and \textit{our model} & Medium & 0.77 (0.30) & $1.15\times
10^{-4} $ $(6.84\times10^{-5})$ \\
$B(l,k)= b (1-|k-l|/(K-1))$ & High & 0.96 (0.10) &
$9.97\times10^{-4} $ $(1.74\times10^{-5})$\\
[6pt]
\multicolumn{4}{@{}c@{}}{\textit{Results for the five-class data
sets}}\\
[4pt]
%$\BB$ model & Risk level & DSC & Estimated risks \\
%[4pt]
\textit{Standard Potts} & Very low & 0.19 (0.21) & $2.45\times
10^{-5} $ $(2.69\times10^{-5})$ \\
$\BB= b I_K$ & Low & 0.25 (0.21) & $7.47\times10^{-5}
$ $(7.01\times10^{-5})$\\
& Medium & 0.32 (0.25) & $1.82\times10^{-4} $ $(1.53\times
10^{-4})$\\
& High & 0.48 (0.32) & $4.15\times10^{-4} $ $(1.81\times
10^{-4})$\\
& Very high & 0.57 (0.25) & $8.47\times10^{-4} $ $(3.95\times
10^{-4})$\\
[4pt]
\textit{Smooth gradation} & Very low & 0.40 (0.24) &
$4.21\times10^{-5} $ $(2.45\times10^{-5})$ \\
$B(l,k)= b (1-|k-l|/(K-1))$ & Low & 0.19 (0.19) & $1.33\times
10^{-4} $ $(1.01\times10^{-4})$ \\
& Medium & 0.14 (0.28) & $2.64\times10^{-4} $ $(2.00\times
10^{-4})$ \\
& High & 0.75 (0.32) & $4.98\times10^{-4} $ $(2.02\times
10^{-5})$ \\
& Very high & 0.89 (0.07) & $1.01\times10^{-3} $ $(6.09\times
10^{-5})$\\
[4pt]
\textit{Our model} & Very low & 0.56 (0.20) &$ 2.07\times
10^{-5} $ $(1.53\times10^{-5})$ \\
$B(l,k)= b [1-|k-l|/2]_+$ & Low & 0.29 (0.17)& $9.62\times
10^{-5} $ $(4.39\times10^{-5})$ \\
& Medium & 0.09 (0.18) & $3.33\times10^{-4} $ $(1.37\times
10^{-4})$ \\
& High & 0.66 (0.38) & $5.57\times10^{-4} $ $(1.05\times
10^{-4})$ \\
& Very high & 0.83 (0.17) & $1.05\times10^{-3} $ $(7.66\times
10^{-5})$ \\
\hline
\end{tabular*}
\end{table}

Overall, we observe that the three strategies are recovering more
easily the high risk regions than the low risk regions. In general,
when classes disappear, they correspond to the regions of lowest risks.
The $S_{\mathrm{tra}}$ strategy performs satisfyingly compared to other
strategies. In particular, the proportions of correctly allocated
hexagons is improved. Also, with a limited amount of computational
resource, the $S_{\mathrm{tra}}$ is more likely to provide satisfying
results with a better exploration of the parameter space. Additional
experiments on the hidden MRF part confirm that both the \textit{smooth
gradation model} and our model improve over the standard Potts model
and suggest that compared to the \textit{smooth gradation model} our
model will be less likely to lose the more problematic lowest risk
classes as illustrated in Figure~\ref{res5class}(f).

%s6 #&#
\section{The bovine spongiform encephalopathy data set}
\label{esb}
\begin{figure}

\includegraphics{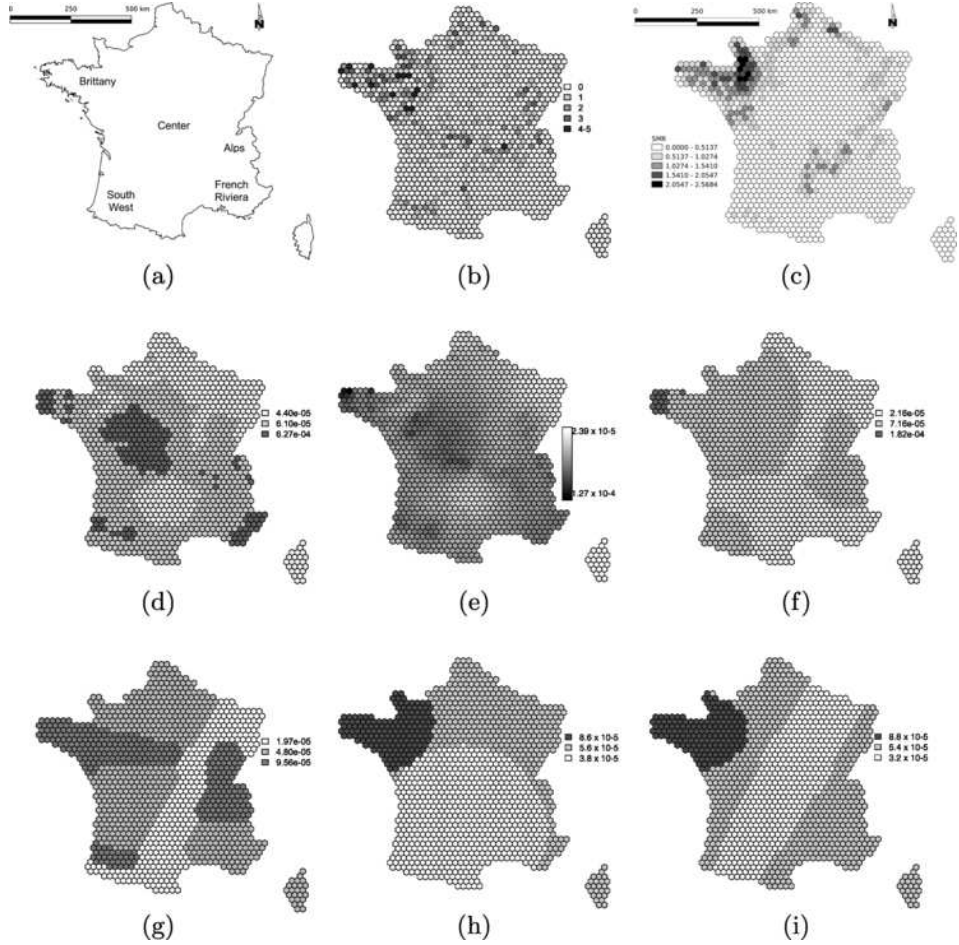}

\caption{BSE data set: \textup{(a)} main France
regions mentioned in the paper, \textup{(b)} BSE cases registered between
July 1, 2001 and December 31, 2005 and \textup{(c)} standard mortality
rates. Estimated risk map \textup{(d)} obtained from continuous risk
values \textup{(e)} using the BYM model. Estimated risk
maps obtained with variational EM for our model \textup{(f)} and the
standard Potts model \textup{(g)}. Estimated risk maps obtained with the
Kulldorff's scan statistics for circular \textup{(h)} and ellipsoidal \textup{(i)}
clusters.} \label{BSEres}
\end{figure}

The Bovine Spon\-gi\-form Encephalopathy (BSE) is a
noncontagious neurodegenerative disease in cattle. This sudden
and unexpected disease %\citep{Anderson96, Ducrot2008}
threatened bovine production in Europe and has been intensively
studied, in particular, for spatial patterns
[\citet{Abrial2005,Allepuz07,Paul07}]. It is transmitted by meat
and bone meal. Since there is no direct transmission and no
vector, spatial analysis is important to understand and explain
the geographical localization of the cases. In our data set, the
numbers of observed cases are available for each
hexagonal geographical unit %s
in France. These cases occurred between July 1, 2001 and December
31, 2005, although at that time meat and bone
meal, the main risk factor,
had been already forbidden for cattle in France. Figure
\ref{BSEres}(b) shows the corresponding observed map,
and Figure \ref{BSEres}(c) the standard
mortality rate.
%2001 and the 31st of December 2005.
%}
We first apply our model initialized with the $S_{\mathrm{tra}}$ strategy
described in Section \ref{stra}. Regarding the number of
classes, the approximated BIC of \citet{ForbesPeyrard03} suggests
to select $K=3$. For comparison, we also consider the BYM model
widely used in epidemiology. Since this model only provides
continuous estimated values for the risk level in each hexagon,
some additional post-processing is required to obtain the mapping
into a prescribed number $K$ of risk levels. Such a mapping can be
obtained by applying some clustering procedure on the estimated
continuous values. A commonly used method for that is the EM
algorithm for Gaussian mixtures. Figure
\ref{BSEres}(d), (f)
shows the mapping obtained
%
%f4 #&#
with the BYM model and our model. The fact that
BYM provides continuous risk values is then not necessarily an
issue for users. As illustrated in a recent paper by
\citet{Hossain2010}, it is natural to consider such models, as they
can have reasonable unusual risk detection behavior and can be
relatively easy to fit. Also, Bayesian approaches represent a whole
family of methods for disease mapping. Cluster detection methods
are another important class of methods. Our data are aggregated,
but we can also consider our data as point data to apply such
cluster detection methods. The objective of cluster detection
methods for point data is to identify, if they exist, the zones in
which the concentration of events is abnormally high, usually
named clusters. To assess the significance of a supposed cluster,
the observed concentration is usually compared with the
concentration observed under the null hypothesis $H_0$ that the
events are sampled independently from the underlying population
density, generally a Poisson distribution in epidemiology (as in the
BYM model and our model). One of the most popular approaches is
the scan statistic adapted to the spatial setting by
\citet{Kulldorff1997}. It relies on the generalized
likelihood-ratio test statistic of $H_0$ against a piecewise
constant density alternative. To apply this method, one needs to
set the family of the possible clusters, for example, all the discs
centered on a point of a predefined grid. The radius of each
circle is generally set to vary continuously from zero to an upper
limit (e.g., less than $50\%$ of the total area). This
predefined shape of the cluster can be an important limitation
since, in the real world, an excess of events may be recorded
along a river, for example. An alternative has been proposed more
recently: \citet{Kulldorff2006} investigated a wide family of
elliptic windows with predetermined shape, angle and center. The
ultimate solution would be to consider all the convex envelopes
including any subset of the events locations. However, this
becomes computationally infeasible when the number of events is
large. The statistical significance of the largest likelihood (for
positive clusters or the lowest likelihood for negative clusters)
is assessed by determining its distribution under the null
hypothesis through Monte Carlo simulation. This method is
implemented in the SaTScan software
[\citet{Satscan2009}, \url{http://www.satscan.org/}]. For the BSE
data set, we then also include the results [Figure
\ref{BSEres}(h), (i)] using the spatial scan statistic for circular
and ellipsoidal clusters. For comparison, we indicate the
corresponding average risk value in each detected cluster.
Positive clusters are indicated using a dark grey color, while
negative clusters are in white.

When comparing the four maps obtained with the
expert knowledge related to the BSE disease in France, the result
from our model
appears to be very satisfying. %Indeed, t
Three regions are clearly delimited and correspond to the regions
expected by the experts and highlighted in
previous works [\citet{Abrial2005b}]. Indeed, in the BSE case it is
known that high risks regions are located in Brittany, in the
center, %Pays de la Loire
in the South--West of France and in the Alps. %Rh{\chr"F4}ne-Alpes
%region.
The localization of these regions is shown in
Figure \ref{BSEres}(a). Geography and topography are not, however,
important explanatory factors for the disease which should rather be
related to local agricultural traditions. In these regions
there is a high density of monogastric species [\citet{Abrial2005}],
for example, pigs and poultry, and meat and bone meal were used to
feed these species [\citet{Paul07}]. It is suspected that the BSE
risk can be explained by cross-contamination with an ingredient
used in poultry or pig feed. Cross-contamination may occur at the
factory, if food chains for monogastrics and ruminants are not
clearly distinct, during the shipment of feed to the farms, or on
the farms, especially on mixed farms with both cattle and pig or
poultry operations [\citet{Abrial2005}]. Our
analysis, detecting coherent risk regions, supports this
hypothesis.

In the BYM map, additional high risk regions are highlighted but
with boundaries that are more doubtful, sometimes including too
few hexagons or including regions\vadjust{\goodbreak} known for low risk. Typically,
the Alps region (known as high risk) is not clearly identified but
merged with the South--East region (known as low risk). Moreover,
the French Riviera appears as a higher risk region than the Alps
and the South--West, although it is known that on this very urban
coast the cattle population is low and no cases
are observed [\citet{Abrial2005}].
%the number of observed cases are low \citep{Abrial2005}.
Our HMRF mapping is in that sense much
more reliable with the French Riviera identified as a low risk
region, as it should. We suspect that this bad feature of BYM may
come from the strength of its spatial prior in the absence of
strong information on the spatial structure in the observed data.
We presume %We suspect
this is the case for the BSE data set so that the
resulting map using BYM may also mainly reflect the prior rather
than the observations.

It can then be seen from the scan statistic maps
in Figure \ref{BSEres}(h), (i) that with this method, among the four
known regions at risk, only Brittany (West) is retrieved as a
positive cluster. The Center, the Alps (East) and the South--West
are not detected as positive clusters. Moreover, these zones
considered at risk are partly included in the negative clusters
detected, that is, highlighted as having a low BSE risk. This
may be related to the fixed shapes of the clusters. Regarding the
higher risk values, they are lower than the ones estimated with
BYM and our model. It is interesting to note a closer similarity
of the ellipsoidal cluster map with the map obtained with the
standard Potts model displayed in Figure \ref{BSEres}(g). In both
maps, the high risk region in the West is too large and a similar
diagonal low risk region is recovered. However, the Potts model is
also able to recover high risk regions in the Alps and
South--West.

When studying diseases, in particular, emerging
diseases, the ability of our method to accurately recover high
risk regions is an essential feature, as it is important to clearly
identify the regions where important and quick decisions have to
be taken to control diseases and to implement prevention measures.
During the BSE epidemic in France, every herd where a case had
been detected was killed. For such a culling protection measure,
with important economical consequences, a better knowledge of risk
regions could be employed to try to limit the culling procedures.
In the case of BSE, the regions highlighted as showing a higher
risk of BSE
%In the BSE case, these regions at risk
through the spatial analysis would have focused
the veterinary services inquiries %in high risk regions
and possibly led to earlier detection of the cross-contamination
factor.

Using the same initialization and computing BIC, we selected also
$K=3$ for both the standard Potts model and \textit{smooth gradation
model} equivalent to our model in this case. Comparing all BIC
values (see Table 3 in the supplemental article
[\citet{supplement}]), the best scores for these two models and
ours are equivalent, but the risk map provided by the \textit{smooth
gradation} and our models [Figure \ref{BSEres}(c)] clearly makes
more sense than the one obtained with the standard Potts model
[Figure \ref{BSEres}(g)]. Once again, we prefer
the method that more accurately recovers high risk regions,
that is, the generalization of the Potts model we propose.

%s7 #&#
\section{Discussion}\label{DISCU}

In this paper we propose an unsupervised method for automatically
classifying geographical units into risk classes,
in order to draw interpretable maps, with
clearly delimited zones. Such risk zones may be
useful to focus posterior disease surveillance, control procedures
and epidemiological studies. To do so, we recast the disease
mapping issue into a clustering task using a discrete hidden
Markov model and Poisson class-dependent distributions. The
designed hidden Markov prior is nonstandard and consists of a
variation of the Potts model where the interaction parameter can
depend on the risk classes. One advantage of our discrete HRMF
modeling is that the classification step is part of the model
instead of being a post-processing step, as in most methods
currently used by animal
epidemiologists. The model parameters are %were
then estimated using an EM algorithm and a mean field
approximation principle. This provides a way to face the
intractability of the standard EM in this spatial context, with a
computationally efficient alternative to more intensive MCMC
procedures. One advantage is then that analysis
is possible on large data sets in a reasonable time. But one
typical limitation is that the uncertainty in both the risk
estimated and classification is likely to be underestimated.
Variational approximation techniques often show competitive
estimations when compared to their MCMC counterparts, but they are
also said to be overly optimistic by underestimating variability.

We then focused on challenges presented by very low risk values
and small numbers of observed cases and population sizes,
as can occur with rare diseases, and as
observed in our real data set regarding BSE in
France. In particular, we addressed the problem of finding good
initial parameter values which can be critical in this context. We
developed a new initialization strategy appropriate for spatial
Poisson mixtures in the event of poorly separated classes, as
encountered in animal disease risk analysis.

Our discrete HMRF-based method provides risk maps more reliable
than the traditional BYM method, with less classification errors
and more clearly delimited at risk zones. Our
generalized Potts model also shows a better ability to recover
risk regions than the standard Potts model. Our experiments show
that our model performs well in determining high risk regions,
both in terms of accurate localization of these regions and
estimation of the associated risk levels. This is an important
point since these high risk regions are of primary interest in
practice when the goal is to eventually impose safety procedures.
The low risk regions are more difficult to delineate, especially
when they are not in areas of large population size.
But their practical interest is less crucial, as
they are essentially used afterward to consider protections
factors. Overall, our experiments suggest that the usual BYM
method, in its simplest version, is not adapted to rare diseases
in very inhomogeneous populations, as it tends to estimate high
risks in regions with very low population.

The solution we propose instead is a flexible model whose
parameters are easy to interpret and to adapt to other situations
involving spatial count data. In particular, the interpretation of
the pair-wise potential functions in terms of neighborhood
interaction allows users to define their own spatial smoothing
depending on the targeted task. As regards disease mapping, we
show that a simple convenient design for the interaction matrix
$\BB$ was provided by our choice of a three nonzero diagonal matrix, which
accounts for smooth risk gradation while producing satisfying
delineations for both high and low risk regions. Also, the
definition of the neighborhood, simply based on geographical
proximity in this paper, can be adapted to the context and
potentially include nonspatial information through some measures
of similarity between sites based on nongeographical features.
Typically, for the BSE example, sites could be set as neighbors if
they share the same animal food provider, information known to be of
major importance regarding the BSE risk
[\citet{Paul07}]. A second example is the possibility to introduce
interactions to account for an ecological gradient, such as wind
dissemination that could be important, for example, for diseases
transmitted by mosquitoes.

In addition, to better understand
the mechanisms underlying the spread of a disease, it is possible
to introduce covariates at various stages of the hierarchy without
changing too much the structure of the model. The use of a mean
field principle for inference generalizes easily in this case and
has the advantage to maintain the model tractability.

Then, the model applies naturally to all kinds of graphical
structures and can therefore adapt easily to integrate temporal
information such as given, for instance, by observations
corresponding to cases for the same area but at different periods
of time. Further investigations for such a spatio-temporal
analysis are planned.


\begin{supplement}%[id=suppA]
\stitle{Supplement to ``Spatial risk mapping for rare disease with
hidden Markov fields and variational EM''}
\slink[doi]{10.1214/13-AOAS629SUPP} %[doi,text={...}] - jei reikia
%suskaldyti doi
\sdatatype{.pdf}
\sfilename{aoas629\_supp.pdf}
\sdescription{Missing appendices, tables
and figures are available in a companion supplemental file.}
\end{supplement}

% imsref loaded by lrinkeviciute, 2013-05-08 14:58:12
% imsref loaded by lrinkeviciute, 2013-05-08 15:38:17
% imsref loaded by lrinkeviciute, 2013-05-08 15:40:39
% imsref loaded by lrinkeviciute, 2013-05-08 15:42:59
% imsref loaded by lrinkeviciute, 2013-05-08 15:44:15
%

\printaddresses

\end{document}